# Induced Ferromagnetism and colossal magnetoresistance by Ir-Doping in $Pr_{1-x}Ca_xMnO_3$

**Sylvie Hébert*, Antoine Maignan, Raymond Frésard, Maryvonne Hervieu, Richard Retoux, Christine Martin and Bernard Raveau**
Laboratoire CRISMAT, UMR 6508 associée au CNRS, ISMRA, 6 Boulevard du Maréchal Juin, 14050 CAEN Cedex, FRANCE.

PACS : 71.30.+h; 75.30.Vm; 75.25.+z

**Abstract**

The doping of the manganese site by iridium (up to 15%) in the small A cation manganites $Pr_{1-x}Ca_xMnO_3$ ($0.4 \leq x \leq 0.8$), has been investigated as a new method to suppress charge-ordering and induce CMR effects. Ir doping leads to ferromagnetism and to insulator to metal transitions, with high transition temperatures reaching 180K and CMR ratio in 7T as large as $10^4$. The efficiency with which iridium induces ferromagnetism and CMR is compared to previous results obtained with other substitutions (Ru, Rh, Ni, Cr…). The ionic radius of the foreign cations and their mixed-valencies are found to be the main parameters governing the ability to collapse the charge-ordered state.

* corresponding author
sylvie.hebert@ismra.fr
fax : 33 (0) 2 31 95 16 00

The double exchange (DE) mechanism [1] which appears between the $Mn^{3+}$ and $Mn^{4+}$ species in the perovskite manganites makes that the Mn-O-Mn bond angles play a prominent role in the appearance of ferromagnetism (FM) and of metallic like properties in these oxides. As a consequence, the size of the A-site cation in the $Ln_{1-x}Ae_xMnO_3$ perovskites (Ln = lanthanide; Ae = alkaline-earth) is a crucial parameter for the existence of colossal magnetoresistance (CMR) in those oxides. For instance, the manganites $Ln_{0.5}Ca_{0.5}MnO_3$ with Ln = Nd, Pr, Sm which are characterized by small A-site cation exhibit a very stable charge ordered CE-type antiferromagnetic (AFM) structure [2-5] and consequently do not show any CMR effect up to 20 teslas [6], in contrast to large A cation manganites such as $Pr_{0.5}Sr_{0.5}MnO_3$, whose A-type AFM structure is destroyed by a magnetic field of ~ 7T [7], leading to a large decrease of the resistivity.

In order to induce CMR properties in small A cations manganites, Mn sites can be doped with magnetic cations such as Cr, Co, Ni [8-9] or Ru [10-11], leading to insulator-metal (IM) transition and to ferromagnetism, even in the absence of a magnetic field. This considerable effect of Mn-site doping is not only due to the destruction of the CO state but originates from the local magnetic (FM or AFM) coupling of the doping element with the surrounding manganese species, leading to the formation of FM domains which extend under the application of a magnetic field [12]. The spectacular effect of ruthenium doping is illustrated by its ability to induce FM and metallicity in electron-doped manganites in contrast to Cr, Co, Ni [12]. The two different possible oxidation states of ruthenium, Ru(V) and Ru(IV), are thought to be responsible for the Ru superiority. In this respect, iridium appears as a potential candidate to induce FM and CMR when substituted in CO manganites. Previous studies of perovskites $Sr_2IrMO_6$ with M = Ta, Nb [13] have indeed shown that Ir(III) is stabilized when the oxides are prepared in air, with a low spin configuration $(t_{2g})^6e_g^0$. Moreover it was recently shown that Ir(IV) exists in hexagonal perovskites $Ba_{1-x}(Cu_xIr_{1-x})O_3$ synthesized in air [14-15], so that fluctuations between Ir(III) and Ir(IV) are also possible in Ir oxides. Such an electronic configuration suggests a great similarity between both $e_g^0$ Ru and Ir cations from the viewpoint of the magnetic properties. We report herein on the doping of the Mn-sites with iridium in the manganites $Pr_{1-x}Ca_xMnO_3$. We show that FM and CMR effect are induced by Ir-doping, with $T_C$ up to 180K.

The synthesis of pure perovskites $Pr_{1-x}Ca_xMn_{1-y}Ir_yO_3$ is made difficult in normal conditions of oxygen pressure, due to the fact that iridium oxide ($IrO_2$) decomposes below 1000°C so that a part of Ir may not be incorporated. The perovskites of the series $Pr_{0.5}Ca_{0.5}$ $Mn_{1-y}Ir_yO_3$ and $Pr_{0.4}Ca_{0.6}Mn_{1-y}Ir_yO_3$ were thus prepared from the mixtures of oxides $Pr_6O_{11}$, CaO, $IrO_2$ and $Mn_2O_3$, according to two different methods depending on the iridium content. For low iridium contents ($y \leq 0.06$), practically pure compounds were synthesized in the form of pellets preheated at 1000°C 12hrs in air and sintered for 12hrs at 1450°C in air. For larger iridium contents ($y > 0.06$), the oxide pellets were heated in silica sealed tubes for 12hrs at 1200°C. The cationic composition of the different samples and especially their homogeneity were checked by energy dispersive analysis. At room temperature, the electron diffraction study showed that the orthorhombic Pnma-type structure remains for $0 \leq y \leq 0.11$.

The ρ(T) curves of the $Pr_{0.5}Ca_{0.5}(Mn_{1-y}Ir_y)O_3$ series (Fig. 1) clearly show the ability of iridium to induce CMR. In the pristine compound, $Pr_{0.5}Ca_{0.5}MnO_3$, as the sample entered in the CO state below $T_{CO}$ ~ 250K, an increase of ρ is observed as shown in the inset of Fig. 1a. By doping with 2% of Ir the room temperature resistivity increases which could be explained by an increase in the charge scattering by the doping element. However, a smoother change at $T_{CO}$ is observed in the Ir doped compound which indicates that the CO establishment starts to be hindered by only 2% doping. According to the ρ(T) curves, the $T_{CO}$ value of $Pr_{0.5}Ca_{0.5}$ $Mn_{0.98}Ir_{0.02}O_3$ is shifted down by about 20K. This reduction of the CO robustness is consistent with the less rapid increase of ρ as T decreases observed below $T_{CO}$. Accordingly, below about 50K, the T dependence of ρ shows a change towards a quasi-T independent behavior. Similarly to the Cr doped CO manganites [16], this plateau at low T can be explained by the coexistence of FM and metallic small regions with the CO-AFM matrix. These FM clusters induced by doping elements are the necessary precursors for the CMR effect, since the clusters growth may allow the percolation threshold to be overcome. The application of a magnetic field of 7T shows that this Ir-doped manganite exhibits CMR, with resistivity ratio RR = $\rho_{0T}/\rho_{7T}$, close to $10^4$ at 80K (Fig. 1a), in contrast to the undoped phase for which a 27T field is required to induce a similar effect. It should be emphasized that although ρ has been dramatically reduced under the application of a magnetic field, the behavior of ρ is not metallic like but exhibits several substructures. The percolation mechanism is reflected in this complex T dependence of ρ. Increasing the iridium content lowers the resistivity significantly at low temperature, leading to an insulator to metal like transition, as shown for y = 0.04 (Fig. 1b) which exhibits a peak at $T_P$ ≈ 150K, and a resistivity of $3.10^{-1}$Ωcm at low temperature; for this doping level the CMR effect is maximum close to the resistivity peak, with RR ~20 under 7T at 140K. For higher iridium contents, the resistivity curves show a different behavior as illustrated for y = 0.10 (Fig. 1c). Although a decrease of ρ is observed in this sample below $T_P$ = 180K, the insulator to metal transition is unachieved since below 150K ρ starts again to increase. Such a different behavior could originate from the lower temperature used for the synthesis in silica tubes (1200°C) compared to other samples prepared in air (1450°C) which may affect the sintering quality and thus lead to a difference in the grain connectivity.

The magnetization curves M(T) measured under 1.45T after a zero field cooling of the $Pr_{0.5}Ca_{0.5}(Mn_{1-y}Ir_y)O_3$ series (Fig. 2) show that Ir-doping induces ferromagnetism, the magnetic moment at 5K increasing rapidly from M = 0.7$\mu_B$ for y = 0.02 up to 2.5$\mu_B$ for y = 0.10. The Curie temperature $T_C$ taken as the inflection points of M(T) curves coincides with the resistivity peak showing that $T_C$ increases with the iridium content up to $T_C$ = 180K for y = 0.10.

These results show that iridium, in spite of its difficulty to enter the perovskite matrix of the manganites, exhibits a similar behavior to ruthenium and chromium when doping $Pr_{0.5}Ca_{0.5}MnO_3$, i.e. like them it induces ferromagnetism and a tendency to metallicity. Doping by iridium leads to a Curie temperatures $T_C$ = 180K which is intermediate between $T_C$'s obtained by Ru and Cr doping, $T_c$ = 230K [10-12] and $T_c$ = 150K [16], respectively.

Finally it is worth pointing out that the effect of Ir-doping decreases with the average size of the A-site cation, as previously observed for other magnetic cations like chromium or ruthenium [8-9]. For instance the oxide $Sm_{0.5}Ca_{0.5}Mn_{0.94}Ir_{0.06}O_3$ exhibits a magnetic moment at 5K of only 0.4$\mu_B$ (Fig. 3), i.e. much smaller than that of $Pr_{0.5}Ca_{0.5}Mn_{0.96}Ir_{0.04}O_3$, of 2$\mu_B$, and is an insulator at 5K without any CMR effect.

In order to compare the efficiency of Ir-doping with that of Ni, Co, Cr and Ru, we have investigated the electron rich manganite $Pr_{0.4}Ca_{0.6}MnO_3$, which exhibits an even more stable CO state than $Pr_{0.5}Ca_{0.5}MnO_3$ if one judges from the charge ordering temperatures, $T_{CO}$ = 270K and $T_{CO}$ = 250K for $Pr_{0.4}Ca_{0.6}$ and $Pr_{0.5}Ca_{0.5}MnO_3$ [2]. It was previously shown that among the magnetic cations reported to be efficient to induce FM in $Pr_{0.5}Ca_{0.5}MnO_3$, i.e. Ni, Co, Cr, Ru, ferromagnetism and metallicity could only be induced in $Pr_{0.4}Ca_{0.6}MnO_3$ by Ru-doping [11]. Synthesizing $Pr_{0.4}Ca_{0.6}Mn_{1-y}Ir_yO_3$ perovskites at high temperature in air (1400°C) lead to inhomogeneous materials, with grains of different cationic compositions for the same sample. Measurements could therefore only be carried out on samples synthesized in sealed tubes at lower temperature (1200°C), which are pure and homogeneous. The ρ(T) curves of the two oxides $Pr_{0.4}Ca_{0.6}Mn_{1-y}Ir_yO_3$ with y = 0.07 and 0.11 (Fig. 4) clearly show that the resistivity, ranging from $10^{-2}\Omega$cm to $5.10^{-1}\Omega$cm in the 4K-350K temperature range, has been decreased by several orders of magnitude for the lowest temperatures with respect to the undoped manganite (inset Fig. 4). The corresponding M(T) curves (Fig. 5) establish without ambiguity the presence of a strong ferromagnetic component with Curie temperatures close to 160K and 180K for y = 0.07 and 0.11, respectively. Thus, ferromagnetism coincides with the appearance of a metallic like behavior (rupture or inflexion on the ρ(T) curves Fig. 4). Note that the value of the magnetic moment at 5K, ranging from 1.8 to 2.0$\mu_B$, is still lower than for the $Pr_{0.4}Ca_{0.6}Mn_{1-y}Ru_yO_3$ series, 2.8$\mu_B$ for y = 0.08 [11], and significantly lower than the theoretical moment of 3.4$\mu_B$. From these observations it appears clearly that although Ir-doping of the CO AFM $Pr_{0.4}Ca_{0.6}MnO_3$ phase promotes the formation of a ferromagnetic phase, its induced amount is not sufficient to achieve a complete insulator to metal transition as is the case for Ru. The oxides $Pr_{0.4}Ca_{0.6}Mn_{1-y}Ir_yO_3$ exhibit negative magnetoresistance, with low resistance ratios under 7T (Fig. 4), due to the rather low values of their resistivity in the absence of a magnetic field.

Measurements for larger $Mn^{4+}$ contents, e.g. for $Pr_{0.2}Ca_{0.8}Mn_{1-y}Ir_yO_3$ series, show that induced ferromagnetism is much smaller, the magnetic moment at 5K reaching only 0.5$\mu_B$ for y = 0.10. Moreover the oxides remain insulators with no CMR. This behavior is fundamentally different from the Ru-doped compounds as illustrated by $Sm_{0.2}Ca_{0.8}Mn_{1-y}Ru_yO_3$ series which still exhibits high magnetic moments, larger than 1.5$\mu_B$ at 5K, with a metal to metal transition, and CMR properties [12].

An ab initio explanation of the above effects is far beyond the capacity of available numerical tools. Instead we resort to qualitative arguments, mostly inspired by Goodenough [20] and previous works on similar systems. In a recent publication [18], we extensively discussed the effect of Rh substitution, which is isovalent to Ir. We therefore assume that the same line of reasoning applies to Ir substitution. Let us briefly summarize it. In order to explain the effect of iridium upon the magnetotransport properties of the manganites, two possibilities can be considered successively, which correspond to the fact that iridium can be either trivalent [13] or tetravalent [14-15]. Let us take as an example the doping of the CE-type AFM structure of $Pr_{0.5}Ca_{0.5}MnO_3$, which consists of FM zig-zag chains of $Mn^{3+}/Mn^{4+}$ cations, coupled antiferromagnetically [2]. The substitution of $Ir^{3+}$ for $Mn^{3+}$ in such a structure would not introduce any ferromagnetism particularly in the low spin $(t_{2g})^6e_g^0$ configuration of $Ir^{3+}$. In contrast, the substitution of one $Ir^{4+}$ cation for one $Mn^{4+}$ cation, should induce ferromagnetism due to the $(t_{2g})^5e_g^0$ S = 1/2 configuration of $Ir^{4+}$.

The substitutions of Mn by various cations does (or does not) lead to an anomalous ferromagnetic metallic state. This gives a hint that the substituting cation participates to the conduction, if any. The second case corresponds to ions which are in $d^0$ configuration (the exception of Fe is discussed below). Here the substitution does not destabilize the insulating state. For these ions, $Mg^{2+}$, $Ti^{4+}$, $Nb^{5+}$ [17], we note that the $e_g$ levels, which are the only ones to allow for a metallic state involving the impurities, are far above the ionization energy. They therefore do not directly contribute to the formation of a metallic state. In contrast, the substitution with various 3d, 4d and 5d cations with partially filled d-shells such as Ni, Co, Cr [8], Ru [10-12], Rh [18], and Ir (this work) results into an anomalous FM state, and therefore calls for a common mechanism. The metallicity of this state turns out to result from the ability of these ions, which are all substantially larger than $Mn^{4+}$, to hinder the establishment of the long-ranged charge-(and orbital) ordered insulating state. There are enhanced valence fluctuations of the type $Ir^{4+} + Mn^{3+} \rightarrow Ir^{3+} + Mn^{4+}$ involving $e_g$ electrons on both cations compared to the non-substituted samples. This mechanism dynamically involves $Ir^{3+}$ in a configuration $(t_{2g})^5 e_g^1$, which is an excited state of $Ir^{3+}$ in the low spin configuration $(t_{2g})^6 e_g^0$. Most probably the valence fluctuations involving $t_{2g}$ electrons are less important: in particular if the ions are ferromagnetically coupled, all three $t_{2g}$ levels are occupied with (say) up spin electrons, which therefore do not move. Another valence fluctuation may result from the motion of the down spin electrons of $Ir^{4+}$. It involves $Mn^{2+}$ configurations, a highly excited state, and therefore occurs scarcely. Finally, transferring the $e_g$ electron of a $Mn^{3+}$ ion to a $t_{2g}$ level of Ir is not expected to be of any relevance given the smallness of the overlap of the orbitals.

The existence of valence fluctuations involving $e_g$ electrons is subject to the condition that a non-filled $e_g$ orbital is available on the substituting atom. This indeed holds for the above ions, in contrast to $Ti^{4+}$ and $Nb^{5+}$, which are in $d^0$ configurations. Obviously such valence fluctuations also take place in the non-substituted samples, even in the CO state. However the symmetry breaking that accompanies the CO transition strongly suggests that one Mn ion mostly couples to two of its six surrounding neighbors. On top, there is experimental evidence in $Pr_{0.5}Sr_{0.41}Ca_{0.09}MnO_3$ that the $Mn^{4+}$ ion is off-centered, with one single "short" $Mn^{4+}$-O distance [19]. This may be visualized as following from having an ion (the $Mn^{4+}$ ion) that is too small to have a sufficient overlap with all its neighbors, and which therefore has to choose one of them as a partner. Without speculating that the $Mn^{4+}$ ion is statically off-centered in the compounds under study, one may nevertheless suspect that substituting $Mn^{4+}$ by, say, substantially larger $Ir^{4+}$, will deeply modify the situation. Indeed there is no reason to imagine the $Ir^{4+}$ ion being off-centered. As a result one may expect that all Ir-O distances will be very similar, in contrast to the Mn-O distances. This higher degree of symmetry may then lead to be having the $d_{x^2-y^2}$ orbital lower in energy than the $d_{3z^2-r^2}$ one. Therefore we conclude that the ratio kinetic energy vs electron-phonon energy will be larger for a $IrO_6$ octahedron than for a $MnO_6$ one. This brings the Ir-substituted system closer to the physics of the double-exchange model, than the non-substituted one. One may therefore expect that the reduction of the influence of the electron-ion coupling will be reflected in the value of the temperature $T_m$ at which the metallic behavior sets in. At this stage it would be tempting to relate $T_m$ to the ionic radius of the substituting atom. Taking as an example Ir substitution, most valence fluctuations will be of the type $Ir^{4+} + Mn^{3+} \leftrightarrow Ir^{3+} + Mn^{4+}$ i.e., the $e_g$ electron of $Mn^{3+}$ is going to $Ir^{4+}$. Since the atomic radius of $Ir^{3+}$ (0.68Å) is larger than the one of $Rh^{3+}$ (0.665Å), one expects a higher $T_m$, which is indeed the case [18]. When trying to extend this reasoning to other elements, one quickly realizes that the atomic radius is far from being the only parameter. Obviously, the spin state and the valency degree of the substituting atom, (5+ for Ru, 2+ for Co, 3+ for Cr) plays an important role as well, which renders a quantitative estimate of $T_m$ difficult. However, regardless of the detailed mechanisms, they all hinder the establishment of long-ranged charge ordering. Two exceptional substituting atoms deserve further comments : Cr and Fe. It is well known that valence fluctuations are strongly suppressed in six-fold coordinated $Cr^{3+}$. In this case, there is a strong antiferromagnetic coupling of Cr to all its neighbours, and ferromagnetism results from a "domino effect" [12]. In contrast to all the above cations, iron substitution produces neither metallicity nor ferromagnetism [17]. Assuming that Fe is 3+ in a high spin configuration, it thus substitutes for $Mn^{3+}$. As far as the ionic radii are concerned, the one of $Fe^{3+}$ and $Mn^{3+}$ are equal. Therefore iron substitution is expected to have little influence on valence fluctuations in the manganites under consideration. There will merely be a small suppression of the charge ordering transition due to disorder effect.

**Concluding remarks**

This study shows that the doping of manganese sites by iridium in $Pr_{1-x}Ca_xMnO_3$ manganites induces ferromagnetism, metallicity and consequently a CMR effect. It is the first time that such an effect is evidenced for a 5d cation. In this respect such properties of iridium can be compared to those recently observed for 3d dopants (Ni, Co, Cr) and 4d dopants (Ru, Rh), which are also effective agents to enhance ferromagnetism and metallic like behavior in manganites. The paradox of iridium deals with the fact that one expects for this cation a great stability of its trivalent state $Ir^{3+}$, whose low spin configuration $(t_{2g})^6e_g^0$ should not induce any ferromagnetism. Thus in comparison to other magnetic cations, one can propose that in the case of Ir doping valence fluctuations involving $e_g$ electrons on both cations of the type $Ir^{4+} + Mn^{3+} \leftrightarrow Ir^{3+} + Mn^{4+}$ are enhanced, bringing the physics of the Ir-doped systems closer to the double exchange model.

**Figure captions :**

Figure 1 : T dependence of the resistivity ρ in 0T and 7T registered upon cooling .

a) $Pr_{0.5}Ca_{0.5}Mn_{0.98}Ir_{0.02}O_3$. Inset of a) : ρ(T) in 0T of $Pr_{0.5}Ca_{0.5}Mn_{1-y}Ir_yO_3$ with y = 0 and y = 0.02.

b) $Pr_{0.5}Ca_{0.5}Mn_{0.96}Ir_{0.04}O_3$.

c) $Pr_{0.5}Ca_{0.5}Mn_{0.9}Ir_{0.1}O_3$.

Figure 2 : T dependence of the magnetization M measured in 1.45T after zero field cooling of $Pr_{0.5}Ca_{0.5}Mn_{1-y}Ir_yO_3$.

Figure 3 : M(T) in 1.45 T of $Pr_{0.5}Ca_{0.5}Mn_{0.96}Ir_{0.04}O_3$ and $Sm_{0.5}Ca_{0.5}Mn_{0.94}Ir_{0.06}O_3$.

Figure 4 : ρ(T) in 0T and 7T of $Pr_{0.4}Ca_{0.6}Mn_{1-y}Ir_yO_3$. Inset : enlargement of ρ(T) in 0T and comparison with y = 0.

Figure 5 : M(T) in 1.45T of $Pr_{0.4}Ca_{0.6}Mn_{1-y}Ir_yO_3$.

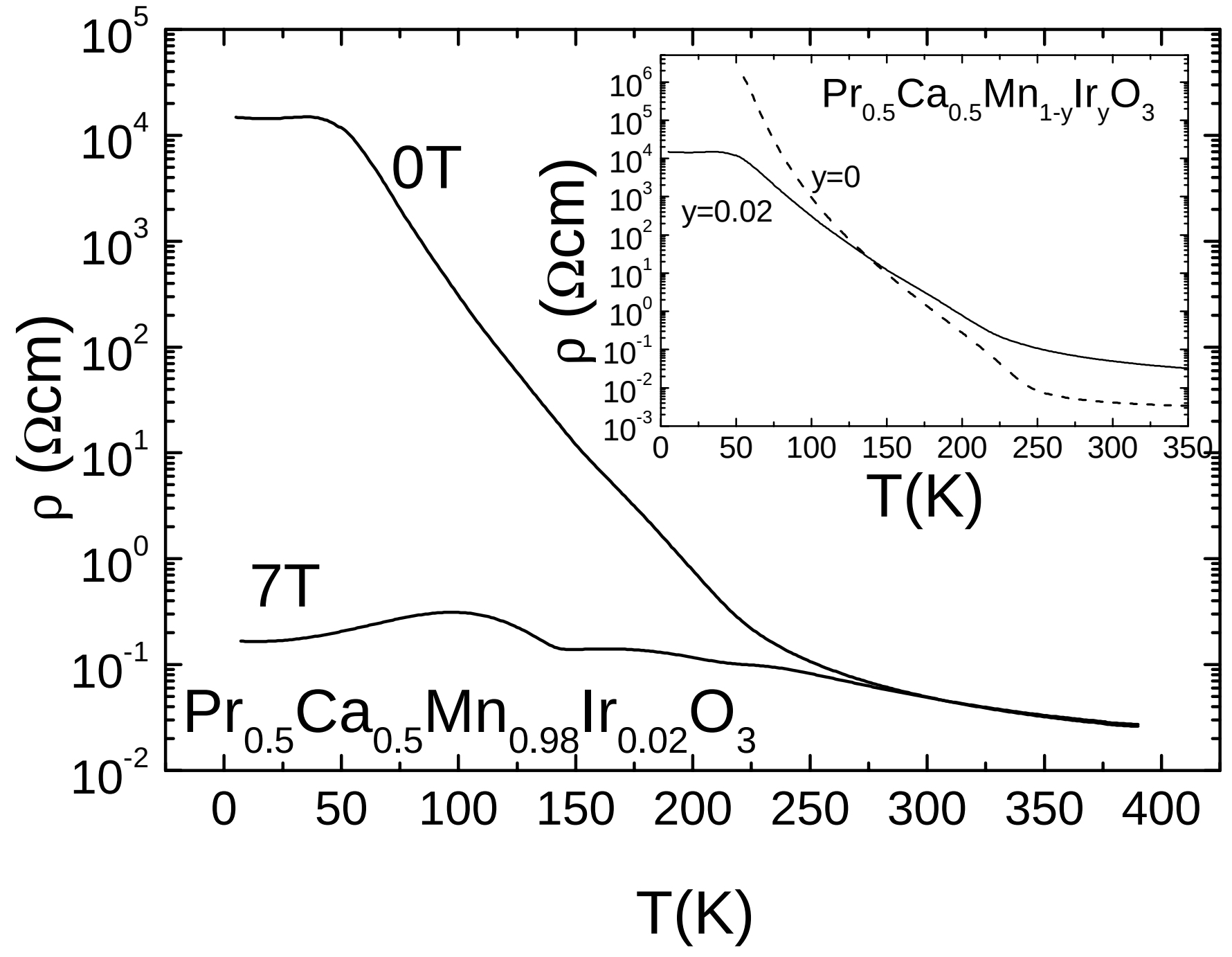

10^5
10^4
10^3
10^2
10^1
10^0
10^-1
10^-2
ρ (Ωcm)
0T
7T
Pr0.5Ca0.5Mn0.98Ir0.02O3
0 50 100 150 200 250 300 350 400
T(K)
Pr0.5Ca0.5Mn1-yIryO3
y=0
y=0.02
ρ (Ωcm)
0 50 100 150 200 250 300 350
T(K)

Fig.1a

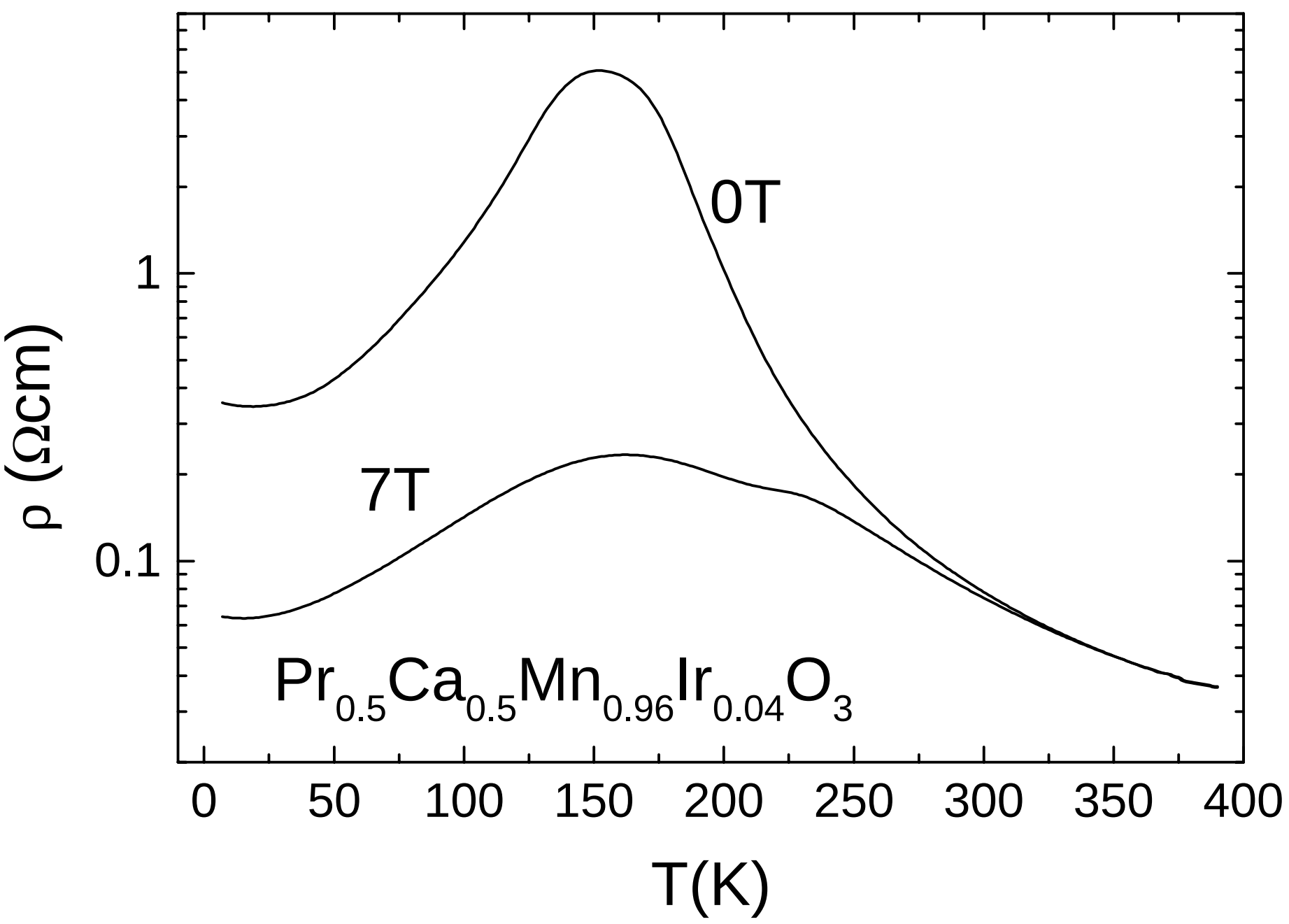

Fig. 1b

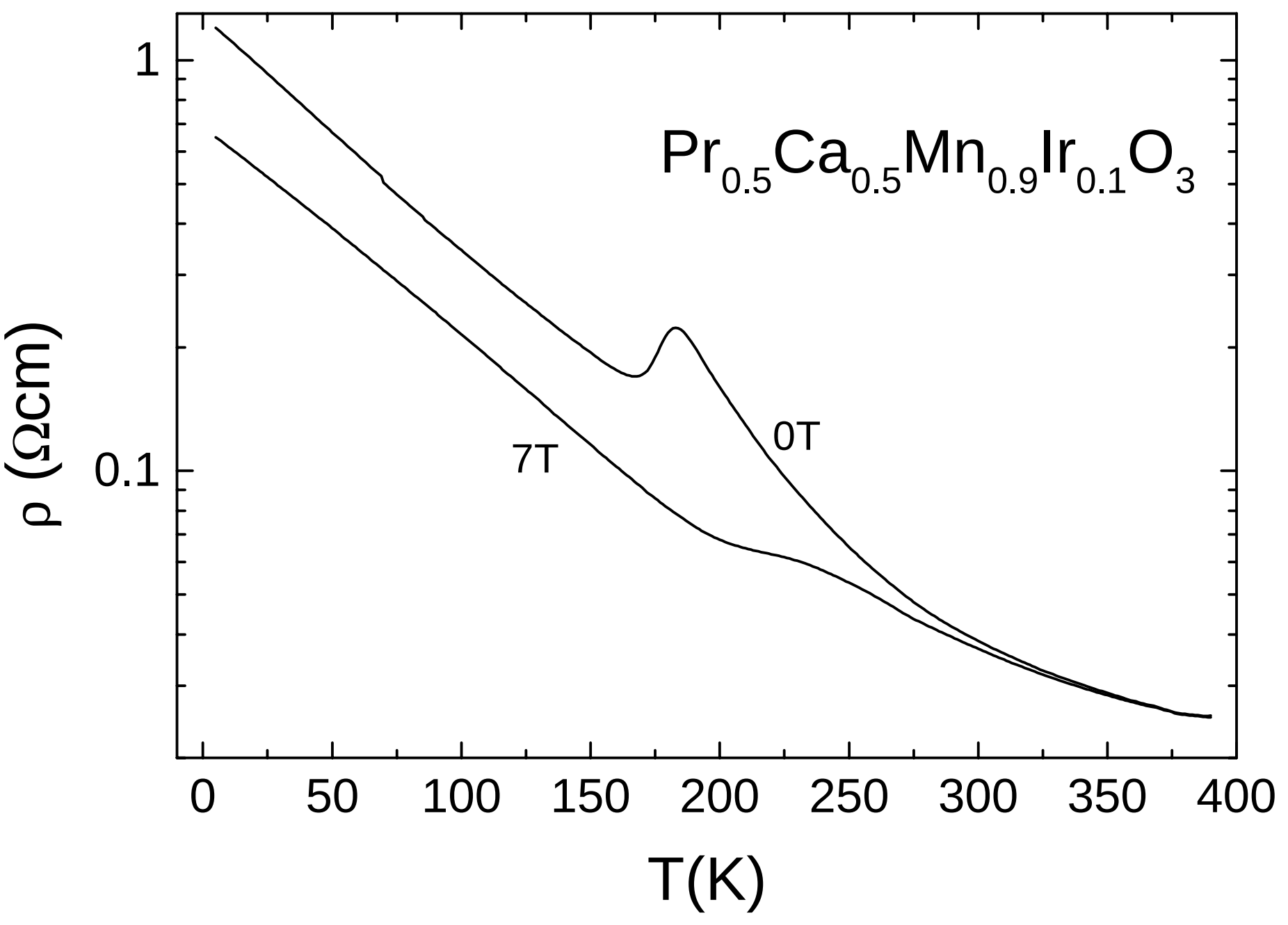

Fig. 1c

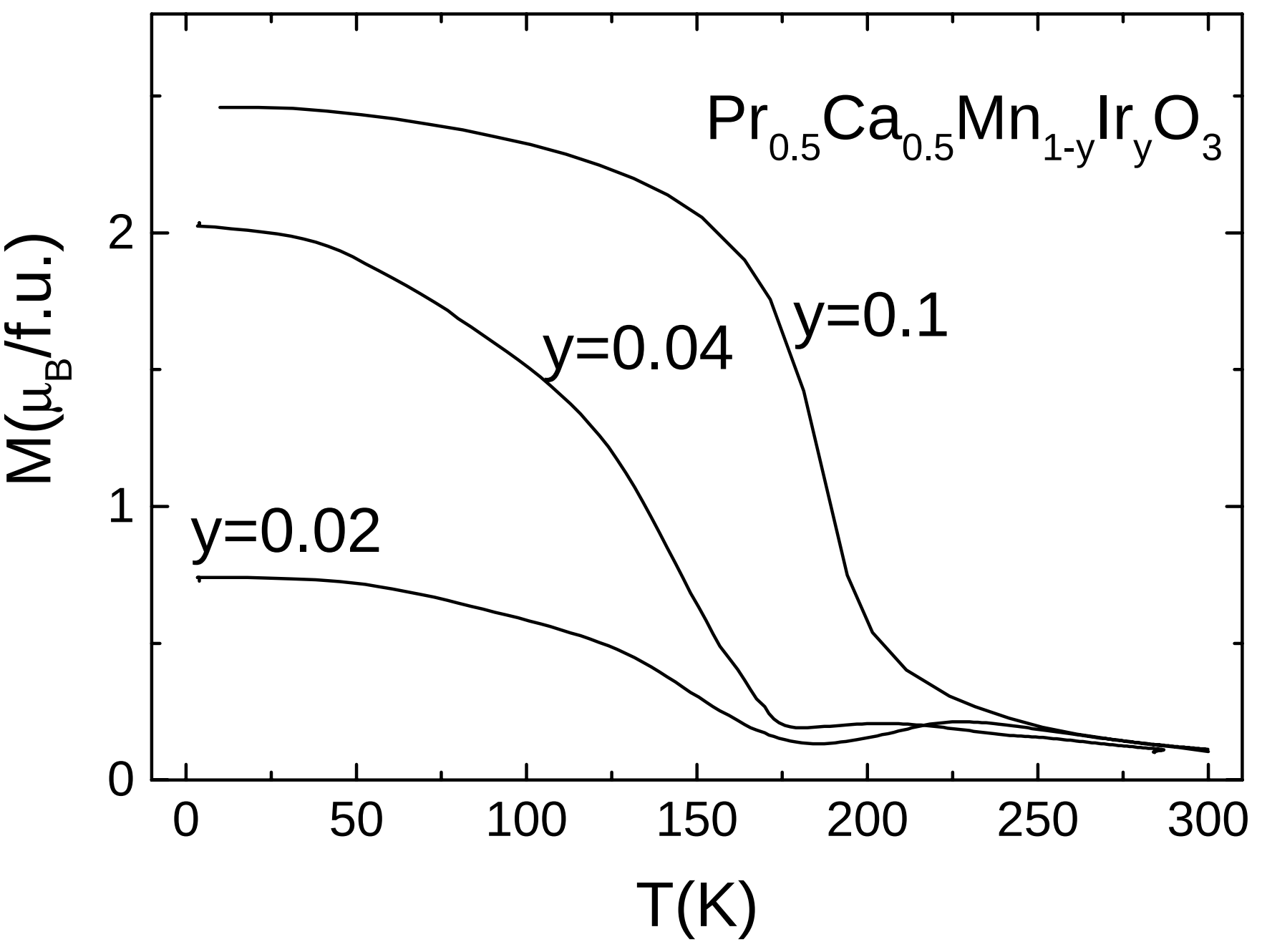

Fig. 2

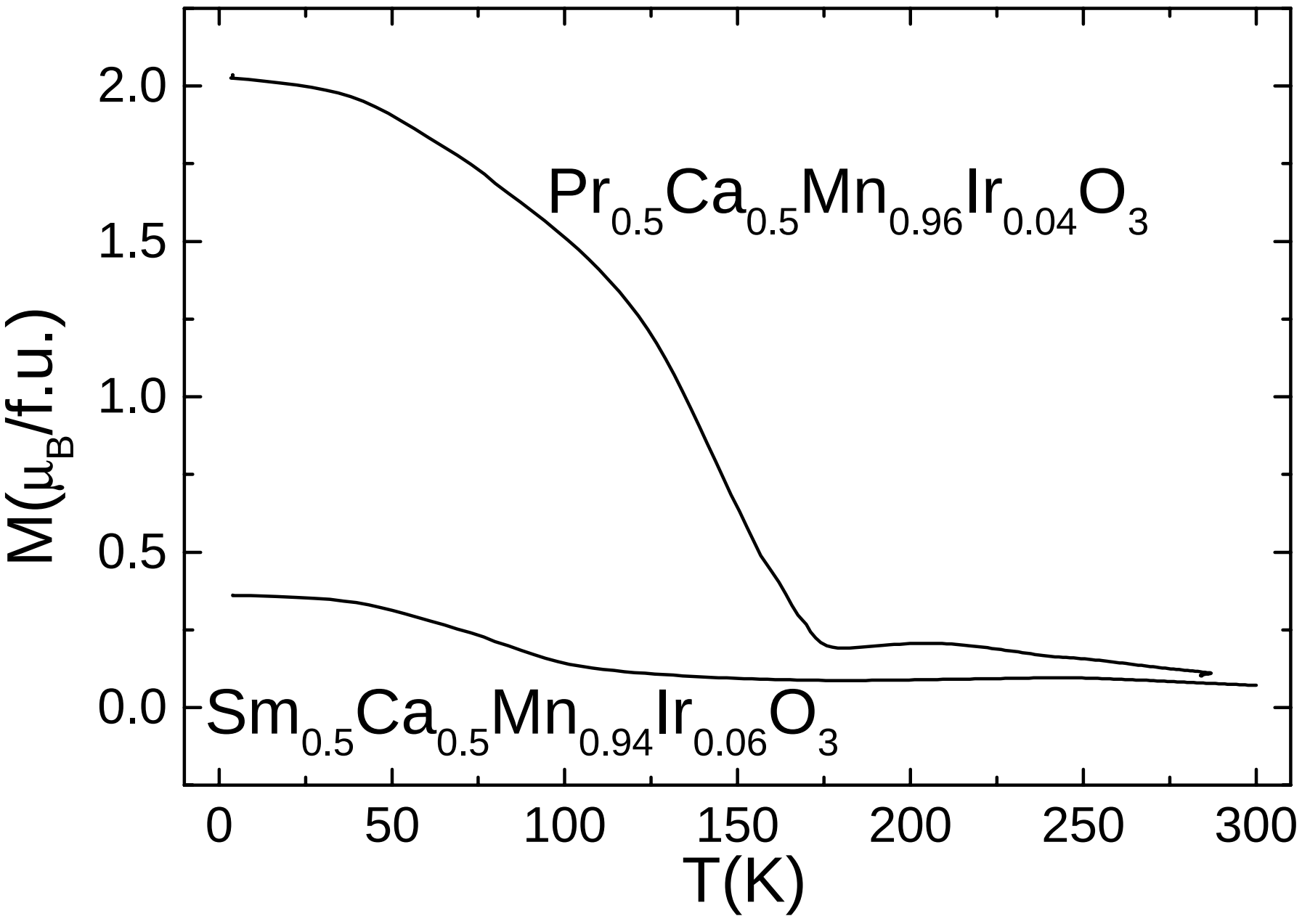

Fig. 3

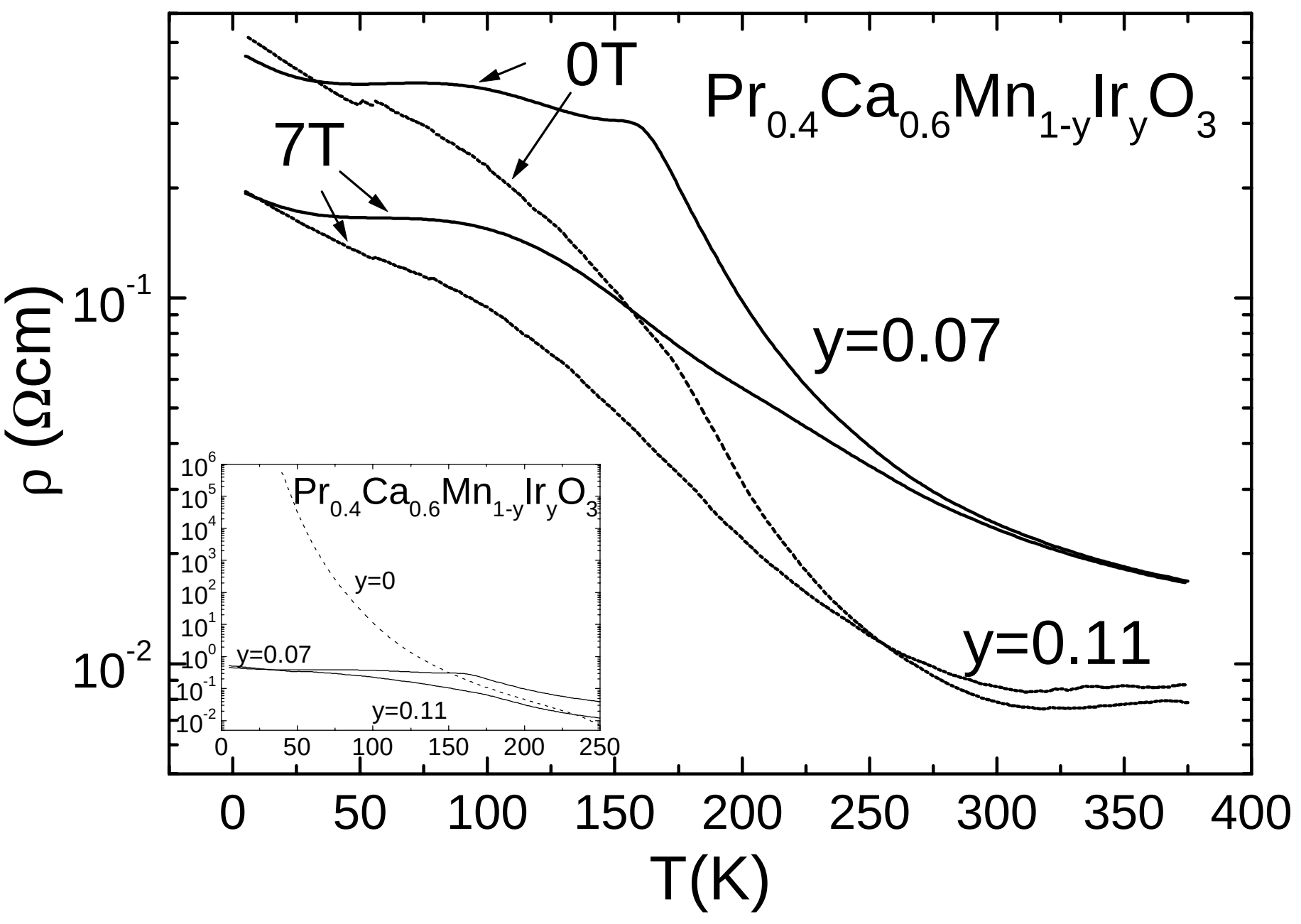

Fig. 4

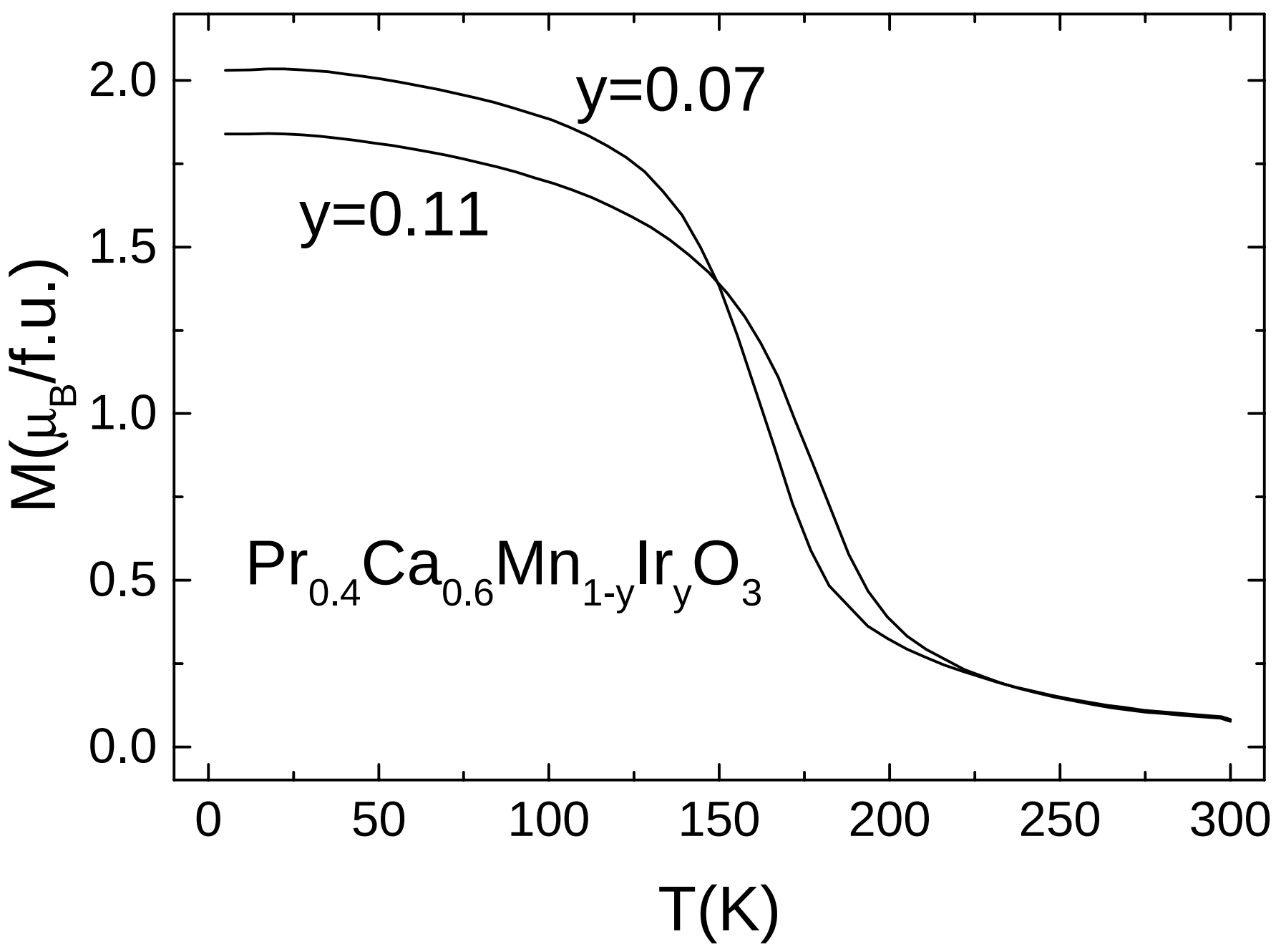

Fig. 5